\documentclass[12pt]{article} 

\begin{document}
\renewcommand{\theequation}{\arabic{section}.\arabic{equation}} 
\newcommand{\beq}{\begin{equation}}  
\newcommand{\eeq}{\end{equation}}    
\newcommand{\bea}{\begin{eqnarray}}  
\newcommand{\eea}{\end{eqnarray}}    
\newcommand{\CN}{{\cal N}}           
\newcommand{\ra}{\rangle}            
\newcommand{\la}{\langle}            
\newcommand{\bb}{|\!|}               
\newcommand{\wt}{\widetilde}         
\newcommand{\non}{\nonumber}         
\newcommand{\rref}[1]{(\ref{#1})}    

\def\sqr#1#2{{\vcenter{\hrule height.#2pt
      \hbox{\vrule width.#2pt height#1pt \kern#1pt
          \vrule width.#2pt}
      \hrule height.#2pt}}}
\newcommand{\ssquare}{{\mathchoice{\sqr34}{\sqr34}{\sqr{2.1}3}{\sqr{1.5}3}}}
\newcommand{\square}{{\mathchoice{\sqr84}{\sqr84}{\sqr{5.0}3}{\sqr{3.5}3}}}
\newfont{\elevenmib}{cmmib10 scaled\magstep1}
\newfont{\cmssbx}{cmssbx10 scaled\magstep3}                     
\newcommand{\preprint}{                                         
            \begin{flushleft}                                   
            \elevenmib Yukawa\, Institute\, Kyoto               
            \end{flushleft}\vspace{-1.3cm}                      
            \begin{flushright}\normalsize  \sf                  
            Preprint YITP-96-53\\ 
            quant-ph./9610022 \\ October 1996       
            \end{flushright}}                                   
\newcommand{\Title}[1]{{\baselineskip=26pt \begin{center}       
            \  \cmssbx #1 \\ \ \\ \end{center}}}                
\newcommand{\Author}{\begin{center}\large \bf                   
            Hong-Chen Fu\footnote[1]{On leave of absence from   
            Institute of Theoretical Physics, Northeast         
            Normal University, Changchun 130024, P.R.China.     
            E-mail: hcfu@yukawa.kyoto-u.ac.jp }\ \              
            and Ryu Sasaki \end{center}}                        
\newcommand{\Address}{\begin{center} \it                        
            Yukawa Institute for Theoretical Physics, Kyoto     
            University,\\ Kyoto 606-01, Japan \end{center}}     
\newcommand{\Accepted}[1]{\begin{center}{\large \sf #1}\\       
            \vspace{1mm}{\small \sf Accepted for Publication}   
            \end{center}}                                       
\baselineskip=20pt
\preprint\bigskip
\Title{ Negative Binomial and Multinomial States:\\  
        probability distributions and coherent states }

\Author
\Address

\vspace{3.3cm}

\begin{abstract}
Following the relationship between probability distribution and 
coherent states, for example the well known Poisson distribution and
the ordinary coherent states and relatively less known one of the 
binomial distribution and the $su(2)$ coherent states,
we propose {\it interpretation} of $su(1,1)$ and $su(r,1)$ coherent states
{\it in terms of probability theory}.
They will be called the {\it negative binomial} ({\it multinomial}\/)
{\it states} which correspond to the {\it negative} binomial (multinomial)
distribution, the non-compact counterpart of the well known binomial
(multinomial) distribution.
Explicit forms of the negative binomial (multinomial) states
are given in terms of various boson representations which are 
naturally related to the probability theory interpretation.
Here we show fruitful interplay of probability theory, group theory 
and quantum theory.
\\ \\

\end{abstract}

\newpage

\section{Introduction}
\setcounter{equation}{0}

It is well known that the photon number distribution of the
ordinary coherent states \cite{Glau,noch,Pere1,Pere} is the Poisson 
distribution, one of the most 
fundamental probability distributions, which  governs random events 
(such as  
radioactive decays) occurring in a  time (space) interval.
As we will show in this paper  the relationship between the coherent 
states in quantum optics and the probability distributions are neither 
coincidental nor superficial but essential.
The main purpose of the present paper is to give  unified 
probabilistic interpretation of the various coherent states.

For the elementary binomial distribution of the probability theory, 
corresponding to the binomial expansion $(1+x)^M=\sum_0^M {M \choose 
n}x^n$, we have $su(2)$ coherent states 
(the `{binomial states' (BS) \cite{stol}) based on the spin $M/2$ 
representation. For the multinomial distributions corresponding to the
multinomial expansion 
\beq(1+x_1+\cdots+x_r)^M=\sum_{n_0+n_1+\cdots+n_r=M} 
{M!\over{n_0!n_1!\cdots 
n_r!}}\, x_1^{n_1}\ldots x_r^{n_r},
\eeq 
we have certain types of $su(r+1)$ coherent states. These coherent
states are known for some time \cite{Gil} but the probabilistic 
interpretation seems new. Let us call them 
{\it multinomial 
states} (MS). They are based on the {\it symmetric} representations 
corresponding to the Young diagram
\beq
\square\square\square \cdots \square\square\ \ M\ {\rm boxes}.
\eeq

In probability theory the non-compact version of the binomial 
distribution is well known and called {\it negative binomial 
distribution}. In this paper the {\it negative binomial states} (NBS) 
of quantised 
radiation field will be introduced in a parallel way as the binomial 
states. It will be shown that they are the well known coherent states of 
$su(1,1)$ algebra
\cite{Pere1,Gil,Gerry,fus3}, the non-compact 
counterpart of the compact $su(2)$ algebra. 
They belong to the discrete series of irreducible 
representations.
Similarly the {\it negative multinomial states} (NMS), the coherent states 
of
$su(r,1)$ algebra belonging to discrete symmetric representations,
will be introduced in terms of the 
{\it negative multinomial distributions}.
It is easy to see that in certain limits these coherent states reduce to 
the ordinary 
coherent states and their tensor products, since the (negative) 
binomial and (negative) multinomial distributions tend to the Poisson 
and multiple Poisson distributions.

\bigbreak

This paper is organised as follows: In section 2 the {\it negative 
binomial states} are introduced directly as a {\it square root} of 
the negative binomial distribution. In other words they are constructed
in such a way that their photon number distribution is the negative 
binomial distribution.
Then these coherent states are shown to have the displacement operator 
forms. Namely, they are created by the action of the unitary 
operators in $SU(1,1)$ acting on certain highest (lowest) weight 
states (``vacuum''). 
In section 3 we relate the inhomogeneous 
representation of $su(1,1)$ suggested by the negative binomial states
to the symmetric two boson realisation.
The two boson formulation provides natural interpretation and more 
explicit 
formulas than those of the formal representation theory of $su(1,1)$.
At the same time this section uncovers some Lie algebraic structures
hidden in the probability distribution.
The physical and statistical properties of the NBS as well as their 
dynamical generation are discussed in some detail in  
our recent publication \cite{fus6}. 
Section 4 deals with the 
generalisation to $su(r,1)$, the negative multinomial states.
One formulation of the negative multinomial states is closely related 
with the Holstein-Primakoff (H-P) \cite{hp} type realisation of 
$su(r,1)$ in terms of 
$r$ (= rank of $su(r,1)$) bosons.
Whereas the comparison with the $r+1$ boson realisation gives 
natural interpretation of various quantities and concepts.
By explicit Lie algebraic calculation which goes quite parallel with 
probability theory, it is shown that the negative multinomial states 
are $su(r,1)$ coherent states belonging to discrete symmetric 
representations. Section 5 is for summary and comments.
Appendix A serves to give general background of the paper, relating 
probability theory, coherent states and Lie algebra theory by taking 
elementary examples such as the ordinary coherent states and the 
binomial states. Appendix B also provides some basic elements like 
quantum mechanical generation of coherent states.
A collection of two level atoms is discussed. It gives a good 
physical example of the binomial states and at the same time it 
provides 
simple interpretation of the H-P realisations as well as the 
relationship with the ordinary coherent states.
Appendix C gives the higher rank generalisation of the results of the 
previous two Appendices. Here we advocate a seemingly ill-recognised 
fact that the multinomial states are coherent states of $su(r+1)$ 
belonging to the symmetric representations.
We stress, here as in the main text, the interplay of probability 
theory, Lie algebra theory and quantum mechanics exemplified in 
various coherent states.
Appendix D gives a short explanation of the negative binomial 
distribution as a distribution of ``waiting time''.
We adopt such notation as to reveal the essential features underlying 
this subject which sometimes results in deviating from the
conventional notation.

\section{Negative Binomial State}
\setcounter{equation}{0}

Let us start with the {\it negative binomial distribution} 
(For an elementary introduction of the  negative binomial 
distribution from  probability theory see Appendix D. For more 
details, see for example, Chap.VI of \cite{Feller})
\begin{equation}
   B_n^-(\eta ;M )={M+n-1 \choose 
   n}\eta^{2n}(1-\eta^2)^M,\quad n=0,1,\ldots,
   \label{nbd}
\end{equation}
in which $0<\eta^2<1$ and $M$ is a positive integer.
This can be rewritten as
\begin{equation}
   ( 1-\eta^2)^{-M}B_n^-(\eta ;M )={-M \choose 
   n}(-\eta^2)^n,\quad n=0,1,\ldots,
   \label{nbd2}
\end{equation}
and it is easy to see that the right hand side corresponds to the 
power series expansion of $(1-\eta^2)^{-M}$, the {\it negative binomial 
expansion}. Thus the normalisation 
\begin{equation}
	 \sum_{n=0}^\infty B_n^-(\eta ;M )=1
 	 \label{norm}
\end{equation}
is obvious. From this it is also easy to see that the 
negative binomial distribution (and later the negative binomial 
states)  can be defined for any 
positive number $M$. In this case we have to interpret
\begin{equation}
	{M+n-1\choose n}={\Gamma(M+n)\over{\Gamma(M)\,n!}}.
	\label{realm}
\end{equation}

Let us introduce the `{\it negative binomial state}' (NBS) by taking 
a `square root' of the negative binomial distribution (\ref{nbd}). To 
be more precise, we follow the analogy {\it Poisson distribution}
$\Leftrightarrow$ {\it coherent state} (for details see Appendix A):
\begin{equation}
 P_n(\alpha)=e^{-\alpha^2}{\alpha^{2n}\over{n!}}\Longleftrightarrow
 |\alpha e^{i\theta}\ra=e^{-\alpha^2/2}\sum_{n=0}^{\infty} {(\alpha 
 e^{i\theta})^n\over\sqrt{n!}}|n\ra,
    \label{chs}
\end{equation}
in which $\alpha>0$.
Namely we define NBS
\begin{equation}
	|\eta e^{i\theta};M\ra^-=(1-\eta^2)^{M/2}\sum_{n=0}^{\infty} 
	\sqrt{M+n-1\choose n}(\eta e^{i\theta})^n\bb n\ra,
	\label{nbs}
\end{equation}
in which  $\{\bb n\rangle\, | n=0,1,\ldots\}$ are the number states of 
an oscillator:
\begin{equation}
	[b,b^\dagger]=1,\quad b\bb0\ra=0,
	\quad \bb n\ra={(b^\dagger)^n\over\sqrt{n!}}\bb0\ra.
	\label{bnum}
\end{equation}
(The reason for using a slightly unconventional notation $\bb n\ra$ 
will become clear in the next section.)
Then the number distribution in the NBS is the negative binomial 
distribution (\ref{nbd}):
\begin{equation}
	|\la n\bb \eta e^{i\theta};M\ra^-|^2
	=(1-\eta^2)^M{M+n-1 \choose n}\eta^{2n}= B_n^-(\eta ;M ).
	\label{numbnbs}
\end{equation}
The condition $0<\eta^2<1$ is necessary for the NBS to be normalisable.
In the next section we will have a geometrical interpretation of the 
same condition as characterising the parameter space (the Poincar\'e 
disk) of the $su(1,1)$ coherent states.

Next let us rewrite (\ref{nbs}) ($\eta_C\equiv\eta e^{i\theta}$)
\begin{equation}
	|\eta_C ;M\ra^-=(1-|\eta_C|^2)^{M\over2}\sum_{n=0}^\infty
	{\sqrt{M(M+1)\cdots(M+n-1)}\over n!} 
	(\eta_C )^n(b^\dagger)^n\bb 0\ra.
	\label{nbs2}
\end{equation}
This can be reexpressed in the exponential form
\begin{equation}
   |\eta_C ;M\ra^-=(1-|\eta_C|^2)^{M\over 2} 
   \exp\left[\eta_C\,{\cal K}_+\right]
   \bb0\rangle,
   \label{disnbs}
\end{equation}
in which
\begin{equation}
   {\cal K}_+=b^{\dagger}\sqrt{M+N} \equiv 
   \sqrt{M+N-1}\,b^{\dagger}. 
   \label{kplu}
\end{equation}
Here use is made of the following identity \cite{fus2}
\begin{equation}
   (b^{\dagger}g(N))^n\bb0\rangle=(b^{\dagger})^n
   g(0)g(1)\cdots g(n-1)\bb0\rangle, \mbox{\ with \ }
   g(N)\equiv \sqrt{M+N},\quad N=b^\dagger b.
   \label{nbinide}
\end{equation}
Eq.(\ref{disnbs}) and (\ref{kplu}) reveal the $su(1,1)$ structure
of NBS since ${\cal K}_+$ and its hermitian conjugate
\begin{equation}
	{\cal K}_-=\sqrt{M+N}\,b\equiv b\sqrt{M+N-1}
	\label{kmin}
\end{equation}
generate the $su(1,1)$ algebra  via  H-P \cite{hp}
realisation of the discrete irreducible representation
\footnote{The generalised case with the real non-integer $M$ gives 
the continuous irreducible representation of the universal covering 
group of $SU(1,1)$.} 
with the Bargman index $M/2$ :
\begin{equation}
	[{\cal K}_+,{\cal K}_-]=-2{\cal K}_0,\quad 
	[{\cal K}_0,{\cal K}_\pm]=\pm {\cal K}_\pm,\quad 
	{\cal K}_0=N+{M\over2},
	\label{su11hp}
\end{equation}
and the ``vacuum'' $\bb0\ra$ is the lowest weight state:
\begin{equation}
	{\cal K}_-\bb0\ra=0,\quad {\cal K}_0\bb0\ra={M\over2}\bb0\ra.
	\label{nbinlow}
\end{equation}

It is easy to see tat  
(\ref{disnbs}) is expressed in the displacement operator form
by using the disentangling theorem for $su(1,1)$ :
\begin{equation}
   |\eta e^{i\theta} ;M\ra^-= 
   \exp\left[\zeta_C\,{\cal K}_+-\zeta_C^*\,{\cal K}_-\right]
   \bb0\rangle,\quad  
   \zeta_C=e^{i\theta}\hbox{arctanh}\,\eta.
   \label{disnbs2}
\end{equation}
In other words the negative binomial states are $su(1,1)$ coherent states
 in the 
definition of \cite{Pere,Pere1,Gil}, although the $su(1,1)$ structure is 
not obvious in the original definition of the binomial state
(\ref{nbs}). It should be remarked that in contrast to the binomial 
states which cover all the coherent states of $su(2)$ the negative 
binomial states give only part of the $su(1,1)$ coherent states.
(There are other types of $su(1,1)$ coherent states: for example 
those which are eigenstates of $K_-$,  \cite{Baru}. )

\bigbreak
It should be remarked that the generating function of the negative
binomial state
\begin{eqnarray}
	G^-(\eta;M;t) & = & \sum_{n=0}^\infty t^nB_n^-(\eta;M),
        \qquad |t|\leq1
	\nonumber  \\
	 & = & {(1-\eta^2)^M\over{(1-\eta^2t)^M}}
	\label{genfun}
\end{eqnarray}
has a succinct ``quantum'' definition
\begin{equation}
	G^-(\eta;M;t)=\la \eta e^{i\theta};M\bb t^N\bb 
        \eta e^{i\theta};M\ra.
	\label{genfun2}
\end{equation}
As is well known in probability theory \cite{Feller} the generating 
function is quite useful for calculating various statistical 
quantities of the negative binomial states \cite{fus6}.

\section{Two Boson Formulation of NBS }
\setcounter{equation}{0}
As with the binomial states discussed in Appendix A and B,
the simplest way to understand the negative binomial states 
algebraically is
to introduce two bosonic oscillators to express the $su(1,1)$
generators as bilinear forms rather than the inhomogeneous forms
as given (\ref{kplu}), (\ref{kmin}) and (\ref{su11hp}).
(We choose the formalism that the oscillators define the ordinary
positive definite Hilbert space but the generators of the algebra 
reflect the non-compactness.)
Let us introduce two bosonic oscillators 
\begin{equation}
	 [a_j,a_k^\dagger]=\delta_{jk},\quad j,k=0,1, 
 	\label{twoosci}
\end{equation}
and the Fock space
\begin{equation}
	a_j|0,0\ra=0,\quad j=0,1,
	\qquad |n_0,n_1\ra=
        {a_0^{\dagger n_0}a_1^{\dagger n_1}\over\sqrt{n_0!n_1!}}
	|0,0\ra.
	\label{twobosfock}
\end{equation}
Define
\begin{equation}
	K_+=a_0^\dagger a_1^\dagger,\quad K_-=a_0a_1,
	\quad K_0={1\over2}(N_0+N_1+1),\quad N_j=a_j^\dagger a_j,
	\label{linsu11}
\end{equation}
which satisfy  $su(1,1)$ algebra
\begin{equation}
	[K_+,K_-]=-2K_0,\quad [K_0,K_\pm]=\pm K_\pm.
	\label{11alge}
\end{equation}
These operators either increase ($K_+$) or decrease ($K_-$) $n_0$ 
and $n_1$ 
simultaneously by 1 or keep them unchanged ($K_0$).
In other words the above Fock space gives a reducible representation
of $su(1,1)$ since the subspaces with different 
$\Delta\equiv n_0-n_1$ are always separated.
So we can restrict it as in the case of the binomial states
\begin{equation}
	\Delta\equiv n_0-n_1=M-1\ge0,
	\label{su11res}
\end{equation}
in which $M$ is a {\it positive integer}
\footnote{For the other sign of r.h.s. we can change the role of 0 
and 1.}
\footnote{In terms of the ``waiting time'' interpretation of the 
negative binomial distribution (see Appendix D) $n_0$ is the total 
number of trials but the last (which is always a {\it success}), 
$n_1$ is
the number of {\it failures} and $M$ is the preset number of {\it 
successes} to be achieved.
}. Thus we arrive at the
discrete representation of $su(1,1)$ with Bargman index $M/2$
\begin{equation}
	|M+n-1,n\ra,\quad n=0,1,\ldots,
	\label{su11linfock}
\end{equation}
with the lowest weight state
\begin{equation}
	K_-|M-1,0\ra=0,\quad K_0|M-1,0\ra={M\over2}|M-1,0\ra.
	\label{su11low}
\end{equation}
Obviously this representation is irreducible.
Since these states are uniquely specified by $n\equiv n_1$, we can 
identify them with the number states defined in the previous section
(\ref{bnum}) together with the ``reduced'' oscillator $b$ and 
$b^\dagger$:
\begin{equation}
	\bb n\ra=|M+n-1,n\ra,\quad n=0,1,\ldots.
	\label{rediden}
\end{equation}
Thus we obtain the H-P representation of $K_\pm$ and $K_0$:
\begin{equation}
	K_+\to{\cal K}_+=b^{\dagger}\sqrt{M+N},\quad
	K_-\to{\cal K}_-=\sqrt{M+N}\,b,\quad
	K_0\to{\cal K}_0=N+{M\over2}.
	\label{kkiden}
\end{equation}
One advantage of the H-P type realisation as above is that it
admits the generalisation to the continuous representation
for non-integer $M$. 

\bigbreak 
The other group theoretical aspects of the negative binomial states 
 are about the same as those in the binomial states.
The physical and statistical properties of the NBS as well as their 
 dynamical generation are discussed in some detail in  
our recent publication \cite{fus6}.
The content of this section, though known in Lie algebra theory, can 
be considered to provide some Lie algebraic backgrounds for 
the probability distribution, which are new to the best of our 
knowledge.

\section{Negative Multinomial States}
\setcounter{equation}{0}
The negative multinomial distribution is
\begin{equation}
	M_{\bf n^\prime}^-(\hbox{\boldmath $\eta$}\,;M)=
	(1-\hbox{\boldmath $\eta$}^2)^M
	{(M+n_1+\cdots+n_r-1)!\over{{\bf 
	n^\prime}!(M-1)!}}\eta_1^{2n_1}\cdots\eta_r^{2n_r},
	\label{negmulnomdis}
\end{equation}
in which $M$ is a positive integer and
\begin{eqnarray}
	{\bf n}&=&(n_0,n_1,\ldots,n_r),\quad 
	{\bf n^\prime}=(n_1,\ldots,n_r)
	\quad \hbox{\boldmath $\eta$}=(\eta_1,\ldots,\eta_r)\in{\bf 
	R}^r,\non\\
	 0&<&\hbox{\boldmath $\eta$}^2=\eta_1^2+\cdots+\eta_r^2<1,
	\quad {\bf n^\prime}!=n_1!\cdots n_r!.
	\label{negmulnot}
\end{eqnarray}
In particular, the negative {\it trinomial distribution} reads
\begin{equation}
	M^-_{n_1,n_2}(\eta_1,\eta_2\,;M)=
	(1-\hbox{\boldmath $\eta$}^2)^M\,
	{(M+n_1+n_2-1)!\over{n_1!n_2!(M-1)!}}\eta_1^{2n_1}\eta_2^{2n_2}.
	\label{negmultridis}
\end{equation}
This can be easily obtained from the negative binomial distribution
$$
 B_n^-(\eta ;M )={M+n-1 \choose 
   n}\eta^{2n}(1-\eta^2)^M,\quad n=0,1,\ldots,
$$
by a binomial expansion
$$
\eta^2=\eta_1^2+\eta_2^2,\quad 
\eta^{2n}=\sum_{n_1+n_2=n}{n!\over{n_1!n_2!}}\eta_1^{2n_1}\eta_2^{2n_2},
$$
and collecting appropriate terms.
By repeating the same thing or by applying a multinomial expansion
we arrive at the general form of the negative multinomial distribution
(\ref{negmulnomdis}). 
As we will see later this procedure also explains the generation of
{\it negative multinomial states}.

\bigbreak
The {\it negative multinomial state} (NMS) is defined by taking a 
``square root'' of the negative multinomial distribution 
(\ref{negmulnomdis}):
\begin{equation}
	|\hbox{\boldmath $\eta$}_C\,;M\ra^-=
	(1-|\hbox{\boldmath $\eta$}_C|^2)^{M/2}\sum_{\bf n^\prime}
	      \sqrt{(M+n_1+\cdots+n_r-1)!\over{{\bf 
	n^\prime}!(M-1)!}}(\eta_{1C})^{n_1}\cdots(\eta_{rC})^{n_r}
	      \bb{\bf n^\prime}\ra,
	\label{negmultist}
\end{equation}
in which the ``reduced'' states  
$\{\bb {\bf n^\prime}\rangle=\bb n_1,\ldots,n_r\ra\, 
| n=0,1,\ldots\}$ are the number states of 
$r$ bosonic oscillators:
\begin{eqnarray}
	[b_j,b_k^\dagger]&=&\delta_{jk},\quad b_j\bb{\bf 0}\ra=0,\quad 
	j=1,\ldots,r 
	\quad \bb{\bf 0}\ra=|0,\ldots,0\ra, \non\\
	\bb{\bf n^\prime}\ra&=&
	{({\bf b}^\dagger)^{\bf n^\prime}\over\sqrt{{\bf n^\prime}!}}\bb{\bf 
	0}\ra,
	\quad 
	({\bf b}^\dagger)^{\bf 
	n^\prime}={b_1^\dagger}^{n_1}\cdots{b_r^\dagger}^{n_r}.
	\label{bmulnum}
\end{eqnarray}
It should be remarked that both negative multinomial distribution
(\ref{negmulnomdis}) and state (\ref{negmultist}) are also well defined 
for $M$ positive real number.

In order to show that the negative multinomial states are the
coherent states of $su(r,1)$, we need to realise the algebra.
Let us first construct $su(r,1)$ generators on the Fock space 
generated by $r+1$ bosonic oscillators:
\begin{eqnarray}
	[a_j,a_k^\dagger]&=&\delta_{jk},\quad a_j|{\bf 0}\ra=0,
	\quad j=0,1,\ldots,r, \quad	|{\bf 0}\ra=|0,0,\ldots,0\ra,\non\\ 
	|{\bf n}\ra&=&{({\bf a}^\dagger)^{\bf n}\over\sqrt{{\bf n}!}}|{\bf 
	0}\ra,\quad 
	({\bf a}^\dagger)^{\bf 
	n}={a_0^\dagger}^{n_0}{a_1^\dagger}^{n_1}\cdots{a_r^\dagger}^{n_r}.
	\label{negmulosci}
\end{eqnarray}
 
Let us  define the $u(r,1)$ generators as bilinears in $a_j$ and 
$a_k^\dagger$: 
\begin{eqnarray}
K_{+j}&=&a_0^\dagger a_j^\dagger,\quad K_{-k}=a_0 a_k,\quad 1\leq 
j,k\leq r,\non\\
K_{jk}&=&a_j^\dagger a_k\quad (j\neq k\neq0),\quad N_j=a_j^\dagger a_j.
\label{sur1lin}
\end{eqnarray}
It is easy to see that they leave the combination
$$\Delta\equiv N_0-(N_1+\cdots+N_r)
$$
invariant and the above Fock space (\ref{negmulosci}) 
is a disjoint sum of subspaces characterised by the value of $\Delta$.
As before, let us impose a constraint
 \begin{equation}
	 \Delta\equiv N_0-(N_1+\cdots+N_r)=M-1\ge0,
 	\label{negmulcon}
 \end{equation}
in which $M$ is a positive integer. Then for fixed $M$ the restricted
space provides an irreducible representation of $u(r,1)$.
It has a lowest weight vector
\begin{equation}
	K_{-j}|M-1,0,\ldots,0\ra=0,\quad j=1,\ldots,r
	\label{nmuklow}
\end{equation}
which is invariant under $su(r)$:
\begin{equation}
	K_{jk}|M-1,0,\ldots,0\ra=N_j|M-1,0,\ldots,0\ra=0,\quad 1\leq j,\ 
	k\leq r.
	\label{nmulinv}
\end{equation}

Let us connect the states $\bb {\bf n^\prime}\ra$ and $|{\bf n}\ra$.
Each state in the above representation is uniquely specified by ${\bf 
n^\prime}$ only and we identify
$$\bb {\bf n^\prime}\ra\equiv|{\bf n}\ra=|n_0,n_1,\ldots,n_r\ra,
\quad n_0=M+n_1+\cdots+n_r-1
$$
or
\begin{equation}
	\bb n_1,\ldots,n_r\ra\equiv |M+n_1+\cdots+n_r-1,n_1,\ldots,n_r\ra,
	\quad \hbox{and} \quad \bb{\bf0}\ra=|M-1,0,\ldots,0\ra.
	\label{nmuliden}
\end{equation}
On these states the $su(r,1)$ generators are expressed 
inhomogeneously:
\begin{eqnarray}
K_{+j}\to{\cal K}_{+j}&=&b_j^\dagger\sqrt{M+N_1+\cdots+N_r-1},
\quad
K_{+j}\to{\cal K}_{+j}=\sqrt{M+N_1+\cdots+N_r-1}\,b_j,\non\\
K_{jk}\to{\cal K}_{jk}&=&b_j^\dagger b_k,\quad {\cal N}_j=N_j.
	\label{k-kcor}
\end{eqnarray}
Note that the invariant subalgebra $u(r)$ is expressed bilinearly.

\bigbreak
It is not difficult to generate the negative multinomial states explicitly 
by applying the $SU(r,1)$ operator on the lowest weight  state 
$\bb{\bf 0}\ra$.
For simplicity and concreteness let us show this for the  $su(2,1)$ case:
\begin{equation}
	|\eta_1e^{i\theta_1}, \eta_2e^{i\theta_2}\,;M\ra^-=
	(1-\hbox{\boldmath $\eta$}^2)^{M\over2}\sum_{n_1,n_2=0}^\infty
	\sqrt{(M+n_1+n_2-1)!\over{n_1!n_2!(M-1)!}}\left(e^{i\theta_1}
        \eta_1\right)^{n_1}
	\left(e^{i\theta_2}\eta_2\right)^{n_2}\bb n_1,n_2\ra.
	\label{negmultridis2}
\end{equation}
For the `{\it negative trinomial state}' we only have to use the 
$su(1,1)$ and $su(2)$
disentangling theorems  for the two  subalgebras spanned by 
$(0,1)$ and $(1,2)$ oscillators, respectively:
\begin{eqnarray}
	& &e^{r(e^{i\theta_1}{\cal K}_{10}-e^{-i\theta_1}{\cal K}_{01})}  
	\label{disen01}  \\
	 & = & \exp[e^{i\theta_1}\tanh 
	 r\,{\cal K}_{10}]\exp[\log(1-\tanh^2r)(N_1+M/2)]
         \exp[-e^{-i\theta_1}\tanh 
	 r\,{\cal K}_{01}],
	\non\\
 & &e^{r^\prime(e^{i\theta_2}{\cal K}_{21}-e^{-i\theta_2}{\cal K}_{12})} 
	\label{disen02} \\
	 & = & \exp[e^{i\theta_2}\tan 
	 r^\prime\,{\cal K}_{21}]\exp[-{1\over2}
	 \log(1+\tan^2r^\prime)(N_1-N_2)]\exp[-e^{-i\theta_2}\tan 
	 r^\prime\,{\cal K}_{12}].
	\non	
\end{eqnarray}
First let us choose $r$ such that
$$
\tanh^2r=\eta_1^2+\eta_2^2,
$$
to obtain
\begin{equation}
	e^{r(e^{i\theta_1}{\cal K}_{10}-e^{-i\theta_1}{\cal K}_{01})}
        \bb0,0\ra=
	(1-\eta_1^2-\eta_2^2)^{M\over2}\sum_{n=0}^\infty
	\sqrt{(M+n-1)!\over{(M-1)!n!}}\,
	\left(e^{i\theta_1}\sqrt{\eta_1^2+\eta_2^2}\right)^n\bb n,0\ra,
	\label{negtwobinst}
\end{equation}
which is a negative binomial state in the $(0,1)$ subspace. Next let us 
choose $r^\prime$ such that
$$
\tan^2r^\prime={\eta_2^2\over{\eta_1^2}},
$$
to obtain
\begin{equation}
	e^{r^\prime(e^{i\theta_2}{\cal K}_{21}-e^{-i\theta_2}{\cal K}_{12})}
	\times e^{r(e^{i\theta_1}{\cal K}_{10}-e^{-i\theta_1}{\cal 
	K}_{01})}\bb0,0\ra=
	|\eta_1e^{i\theta_1}, \eta_2e^{i\theta_2}\,;M\ra^-.
	\label{negtrinomgen}
\end{equation}
Thus we have shown that the negative trinomial states are the coherent 
states of $su(2,1)$ belonging to discrete symmetric representations.
Note the parallelism with the negative trinomial distribution at the 
beginning of this section.
The generalisation to higher rank cases is straightforward.
One has to apply first $su(1,1)$ disentangling theorem and $su(2)$
disentangling theorems in the following 
sequence of $su(2)$ algebras spanned by
$(1,2)$, $(2,3)$, $\ldots$, $(r-1,r)$ oscillators.

\bigbreak
Before concluding this section let us remark that the generalisation 
of the discussion (\ref{negmultridis2})--(\ref{negtrinomgen}) to the 
coherent states of $su(r,s)$ is rather straightforward.

\section{Summary and Comments}
\setcounter{equation}{0}
Stimulated by the well known fact that the photon number 
distribution of the ordinary coherent states is Poissonian,
we have constructed quantum mechanical states which have the other
well known probability distributions such as the binomial, multinomial,
negative binomial and negative multinomial distributions 
as their particle number distributions.
They turn out to be the coherent states of the well known 
Lie algebras of $su(2)$, $su(r+1)$, $su(1,1)$ and $su(r,1)$,
respectively, belonging to certain symmetric representations.
Interpretation of these coherent states in terms of probability 
theory is obtained and it is quite useful.
At the same time Lie algebraic structure of these most fundamental
probability distributions is revealed.

The results of the present paper provoke many questions, to
most of which we do not have answers yet.
For example: What about the coherent states of $su(r+1)$ ($su(r,1)$)
belonging to the other representations? 
\footnote{Some coherent states can be constructed in terms of fermion 
oscillators \cite{noch,Pere}.}
Are they also characterised by some probability distributions?
The same question for the other Lie algebras, in particular the 
exceptional ones.
Or the affine Lie algebras and other infinite dimensional algebras
like the Virasoro and the $w$-algebras\ldots?

\section*{Acknowledgments}
We thank K.\,Fujii for useful discussion.
This work is supported partially by the grant-in-aid for
Scientific Research, Priority Area 231 ``Infinite Analysis'',
Japan Ministry of Education. H.\,C.\,F is grateful to Japan
Society for Promotion of Science (JSPS) for the fellowship.
He is also supported in part by the National Science
Foundation of China.

\section*{Appendix A\ \ Binomial States}
\setcounter{equation}{0}
\renewcommand{\theequation}{A.\arabic{equation}} 

In the Appendix A-C, we reformulate the mathematical theory of 
coherent states for $su(2)$ and $su(r+1)$ algebras.
Most of the results are known in one way or another but we believe that
the elementary exposition and the resulting explicit and concrete 
formulas and the emphasis on the connection with probability 
distributions are helpful and useful for most readers.
It is also hoped that the comparison and the contrast with the 
compact cases will provide deeper understanding of the non-compact 
cases treated in the main sections.

We follow the schematic path
\begin{equation}
	{\rm Probability\ Distribution}\Longleftrightarrow
        {\rm Coherent\ States}
	\label{scheme}
\end{equation}
by imitating the well known example of the Poisson distribution
 \begin{equation}
	 P_n(\alpha)=e^{-\alpha^2}{\alpha^{2n}\over{n!}},\quad n=0,1,2,\ldots.
	 	\label{Apois}
 \end{equation}
 Let us introduce the ``probability amplitude'' by taking its `square 
 root'%
 \footnote{Apart from the well known example Klein-Gordon eq. 
 $\Rightarrow$ Dirac eq., let us mention that the creation and 
 annihilation operators $a^\dagger, a$ are also `square roots' of the
 oscillator Hamiltonian. In all these cases, including the probability 
 amplitudes, the `square roots' are complex in spite of the reality 
 (hermiticity) of the original objects.}
 \begin{equation}
	 |\alpha e^{i\theta}\ra=e^{-\alpha^2/2}\sum_{n=0}^\infty {(\alpha 
	 e^{i\theta})^n\over\sqrt{n!}}|n\ra,
 	\label{cohest}
 \end{equation}
 in which $\{|n\ra |n=0,1,\ldots\}$ are the number states of the 
 ordinary oscillator
 \begin{equation}
 	[a,a^\dagger]=1,\quad a|0\ra=0,\quad 
	 |n\ra={(a^\dagger)^n\over\sqrt{n!}}|0\ra.
 	\label{oscil}
 \end{equation}
 The origin of the additional phase factor $e^{i\theta}$ is obvious,
 $\alpha^2=\alpha e^{i\theta}\alpha e^{-i\theta}$.
 By using the last formula of (\ref{oscil}) we can rewrite 
 (\ref{cohest}) as ($\alpha_C\equiv\alpha e^{i\theta}$)
 \begin{equation}
 	|\alpha_C\ra=e^{-|\alpha_C|^2/2}\sum_{n=0}^\infty
	 {(\alpha_Ca^\dagger)^n\over{n!}}|0\ra
	 =e^{-|\alpha_C|^2/2}e^{\alpha_Ca^\dagger}|0\ra
	 =\exp\left[{\alpha_Ca^\dagger-\alpha_C^* 
	 a}\right]|0\ra.
 	\label{cohedisp}
 \end{equation}
 At the last step use is made of the Baker-Campbell-Hausdorff formula.
 Eq.(\ref{cohedisp}) tells that the parameter space is the ordinary  
 complex plane ${\bf C}$, which is a coset space
 \begin{equation}
	 {\bf C}={\rm Heisenberg\hbox{-}Weyl\ Group}/U(1),
 	\label{coheparam}
 \end{equation}
in which the   Heisenberg-Weyl Group is generated by $a, a^\dagger$ 
and the identity operator. The stability subgroup $U(1)$ is just the 
group of complex numbers of unit modulus, $U(1)=\{e^{i\theta}\,| 
\theta: {\rm real}\}$.

The binomial distribution
\begin{equation}
	B_n(\eta;M)={M\choose n}\eta^{2n}(1-\eta^2)^{M-n},\quad 
	n=0,1,\ldots,M,
	\label{bindis}
\end{equation}
is a well known elementary probability distribution related with 
binomial expansion%
\footnote{Note that $e^{\alpha^2}P_n(\alpha)={\alpha^{2n}\over{n!}}$ 
is obtained by a power series expansion of $e^{\alpha^2}$.}
\begin{equation}
	1=1^M=(1-x+x)^M=\sum_{n=0}^M{M\choose n} x^n(1-x)^{M-n}.
	\label{binexp}
\end{equation}
This gives the probability of $n$ `successes' among $M$ times 
repeated Bernoulli's trials with the {\it success} probability $0<\eta^2<1$.
The associated `probability amplitude' is 
\begin{equation}
	|\eta e^{i\theta};M\ra=\sum_{n=0}^M\sqrt{M\choose n}\, (\eta 
	e^{i\theta})^n(1-\eta^2)^{M-n\over2}\bb n\ra,
	\label{binst}
\end{equation}
in which $\{\bb n\ra\ |n=0,1,\ldots\}$ are the number states of the 
 oscillator
 \begin{equation}
 	[b,b^\dagger]=1,\quad b\bb0\ra=0,\quad 
	 \bb n\ra={(b^\dagger)^n\over\sqrt{n!}}\,\bb0\ra.
 	\label{oscil2}
 \end{equation}
 (The reason for using a different oscillator $b, b^\dagger$ from the 
 above coherent state one $a, a^\dagger$ and the slightly 
unconventional notation 
 $\bb n\ra$ will become clear in Appendix B.)
Let us call the state (\ref{binst}) the `{\it binomial state}' (BS) 
\cite{stol,lee,barr}.
At first glance one might be tempted to give a phase  to the 
second factor
\begin{equation}
	(\sqrt{1-\eta^2} e^{i\theta_2})^{M-n}.
	\label{addphas}
\end{equation}
But this is unnecessary since it is decomposed to an overall phase 
$e^{iM\theta_2}$ (which is immaterial) and $e^{-in\theta_2}$ which can 
be absorbed by the redefinition of $\theta$, $\theta\to\theta-\theta_2$.

Next let us rewrite (\ref{binst}) ($\eta_C\equiv\eta e^{i\theta}$)
\begin{equation}
	|\eta_C;M\ra=(1-|\eta_C|^2)^{M\over2}\sum_{n=0}^M
	{\sqrt{M(M-1)\cdots(M-n+1)}\over n!} \left({\eta_C 
	\over\sqrt{1-|\eta_C|^2}}\right)^n(b^\dagger)^n\bb 0\ra.
	\label{binst2}
\end{equation}
Then, by making use of the following identity \cite{fus2}
\begin{equation}
   (b^{\dagger}g(N))^n\bb0\rangle=(b^{\dagger})^n
   g(0)g(1)\cdots g(n-1)\bb0\rangle, \mbox{\ with \ }
   g(N)\equiv \sqrt{M-N},\quad N=b^\dagger b,
   \label{binide}
\end{equation}
we can write (\ref{binst2}) in the exponential form
\begin{equation}
   |\eta_C;M\ra=(1-|\eta_C|^2)^{M\over 2} 
   \exp\left[{\eta_C\over\sqrt{1-|\eta_C|^2}}\,{\cal J}_+\right]
   \bb0\rangle.
   \label{disbin}
\end{equation}
Here ${\cal J}_+$ 
\begin{equation}
   {\cal J}_+=b^{\dagger}\sqrt{M-N} \equiv 
   \sqrt{M-N+1}\,b^{\dagger} 
   \label{jplu}
\end{equation}
together with its hermitian conjugate
\begin{equation}
   {\cal J}_-=\sqrt{M-N}\,b \equiv 
   b\,\sqrt{M-N+1}\,
   \label{jmin}
\end{equation}
generate the $su(2)$ algebra  via  H-P \cite{hp}
realisation in the spin $M/2$ representation:
\begin{equation}
	[{\cal J}_+,{\cal J}_-]=2{\cal J}_0,\quad 
	[{\cal J}_0,{\cal J}_\pm]=\pm {\cal J}_\pm,\quad 
	{\cal J}_0=N-{M\over2},
	\label{su2hp}
\end{equation}
and the `vacuum' $\bb0\ra$ is the lowest weight state
\begin{equation}
	{\cal J}_-\bb0\ra=0,\quad {\cal J}_0\bb0\ra=-{M\over2}\bb0\ra.
	\label{binlow}
\end{equation}

By using the disentangling theorem for $su(2)$ we can rewrite 
(\ref{disbin}) as
\begin{equation}
   |\eta e^{i\theta};M\ra= 
   \exp\left[\zeta_C\,{\cal J}_+-\zeta_C^*\,{\cal J}_-\right]
   \bb0\rangle,\quad  
   \zeta_C=e^{i\theta}\arctan\left({\eta \over\sqrt{1-\eta^2}}\right).
   \label{disbin2}
\end{equation}
In other words the binomial states are $su(2)$ coherent states in the 
definition of \cite{Pere1,Pere,Gil}, although the $su(2)$ structure is 
not obvious in the original definition of the binomial state
(\ref{binst}). 
Since all the irreducible representations of $su(2)$ are exhausted by
the representations (\ref{oscil2})-(\ref{binlow}) for all non-negative 
integer values of $M$, the binomial states give all the $su(2)$ 
coherent states.

Before closing this Appendix, let us recall the fact that the 
binomial distribution tends to the Poisson distribution in a certain 
limit. Let $M\to\infty$, $\eta\to0$ in such a way that the average 
value $\la n\ra$ is fixed:  $\la n\ra=\eta^2M=\alpha^2$. Then for 
finite $n$
\begin{equation}
	B_n(\eta ;M)\to {\alpha^{2n}\over{n!}} 
	e^{-\alpha^2}=P_n(\alpha).
	\label{Poilim}
\end{equation}
(Of course there are many other ways of showing this, e.g. in terms 
of the generating functions of these distributions.)
We have the corresponding limit at the level of the ``probability 
amplitude'' (\ref{binst})
\begin{equation}
	|\eta e^{i\theta};M\ra \to e^{-\alpha^2/2}\sum_{n=0}^\infty
     {(\alpha e^{i\theta})^n\over{\sqrt{n!}}}\bb n\ra,
	\label{Poilimst}
\end{equation}
namely the binomial state tends to the ordinary coherent state.
This limit can also be visualised as a contraction of $su(2)$ 
(\ref{jplu}),(\ref{jmin}) into
the Heisenberg-Weyl algebra :
\begin{equation}
	\eta {\cal J}_+\to \alpha b^\dagger,\quad 
        \eta {\cal J}_-\to \alpha b.
	\label{su2cont}
\end{equation}
Thus (\ref{disbin2}) tends to
\begin{equation}
	|\eta e^{i\theta};M\ra\to
	   \exp\left[\alpha e^{i\theta}\,b^\dagger-\alpha 
           e^{-i\theta}\,b\right]
	   \bb0\rangle.\quad 
	\label{Poilimst2}
\end{equation}

In Appendix B we will discuss the physical problem of dynamical 
generation of BS starting from certain Hamiltonian.
This, in turn, will provide a mathematical framework in which
(i) $su(2)$ structure is more visible and,  (ii) generalisation to the 
coherent states of $su(r+1)$ algebra, the `{\it multinomial states}', 
is straightforward.

\section*{Appendix B\ \ Binomial States: Two Boson Formulation}
\setcounter{equation}{0}
\renewcommand{\theequation}{B.\arabic{equation}} 

In order to discuss the generation of the binomial states,
let us recapitulate the process of the physical generation of the 
ordinary coherent states, for comparison.
This is an oversimplified model retaining only the most essential
features of the coherent states.
We focus on one particular mode of the photon since the system is 
decomposed into a sum of such subsystems:
\begin{equation}
	H=H_0+H_1,\quad H_0=\omega a^\dagger a,\quad 
	H_1=j(t)\,a^\dagger+j(t)^*\,a,
	\label{cohham}
\end{equation}
in which $a^\dagger,a$ are the creation and annihilation operators of 
the photon and $j(t)$ is the classical current (with complex phase).
The state vector in the interaction picture $|\psi(t)\ra_I$ obeys the 
equation of motion
\begin{eqnarray}
	 i{d\over{dt}}|\psi(t)\ra_I&=& {\cal H}_I(t)|\psi(t)\ra_I,
	\nonumber  \\
	{\cal H}_1(t) &=&e^{iH_0t}H_1e^{-iH_0t}=j(t)e^{i\omega t}\,
        a^\dagger
        +j(t)^*e^{-i\omega t}\,a.
	\label{cohint}
\end{eqnarray}
Let us suppose that the system is in the `vacuum' $|0\ra$ at $t=0$. 
Then we obtain
\begin{eqnarray}
	|\psi(t)\ra_I &= &T\,e^{-i\int_0^t{\cal H}_1(t^\prime)dt^\prime}
	|0\ra, 
	\nonumber  \\
	 &=&e^{i\Omega(t)}e^{(\alpha(t)\,a^\dagger-\alpha(t)^*\,a)}|0\ra,
	 \nonumber  \\
	 \alpha(t)&=&-i\int_0^tj(t^\prime)e^{i\omega t^\prime}dt^\prime,
	\label{cogensol}
\end{eqnarray}
in which $T$ is the time-ordering operator and $\Omega(t)$ is a
 calculable function giving the immaterial overall phase. 
 
 \bigbreak
 For the {\it binomial states} let us consider a slightly different 
 model consisting of a number of identical {\it two level atoms} (bosons).
 Let us also assume that the space extension of the system is not big 
 compared with the wavelength of the photon corresponding to the 
 energy gap and that the interactions between different atoms are 
 negligible.
 The system can be described by the ``spin'' operators
 \begin{eqnarray}
 	H_B&=&H_{B0}+H_{B1},\nonumber \\
 	H_{B0}&=&\epsilon\sum_{j=1}^M\sigma_0^{(j)},\nonumber \\
	 H_{B1}&=&\sum_{j=1}^M\left(\lambda(t)\sigma_+^{(j)}
	     +\lambda(t)^*\sigma_-^{(j)}\right),
 	\label{binham}
 \end{eqnarray}
 in which $M$ is the number of the two level atoms.
 As is well known \cite{Dira} a collection of identical particles can 
 also be described by oscillators corresponding to each energy 
 eigenstate.
 Let us denote the lower (upper) state and the corresponding 
 oscillator by 0 (1) \footnote{
 In quantum optics situations one may call 0 `wiggler' photon and 1 
 `laser' photon \cite{lee}.}:
 \begin{equation}
	 [a_j,a_k^\dagger]=\delta_{jk},\quad j,k=0,1. 
 	\label{twoosci2}
 \end{equation}
 Then the above Hamiltonian is equivalent to
  \begin{eqnarray}
 	H_B^\prime&=&H_{B0}^\prime+H_{B1}^\prime,\nonumber \\
 	H_{B0}^\prime&=&\omega_1 N_1+\omega_0N_0,\quad N_j=a_j^\dagger 
 	a_j,\quad j=0,1,
 	\quad \omega_1-\omega_0=\epsilon\nonumber \\
	 H_{B1}^\prime&=&\mu(t)a_1^\dagger a_0
	     +\mu(t)^*a_0^\dagger a_1,
 	\label{binham2}
 \end{eqnarray}
and its Fock space is
\begin{equation}
	|n_0,n_1\ra,\qquad n_0+n_1=M. 
	\label{binfock}
\end{equation}
Now the $su(2)$ structure is obvious, since
\begin{equation}
	J_+=a_1^\dagger a_0,\quad J_-=a_0^\dagger a_1,\quad 
	J_0={1\over2}(N_1-N_0)
	\label{binsu2}
\end{equation}
generate an $su(2)$ algebra and the Fock space (\ref{binfock}) gives
the $M+1$ dimensional (spin $M/2$) irreducible representation 
corresponding to the Young diagram
\beq
\square\square\square \cdots \square\square\ \ M\ {\rm boxes}
\eeq
with the lowest weight state
\begin{equation}
	J_-|M,0\ra=0,\quad J_0|M,0\ra=-{M\over2}|M,0\ra.
	\label{binlow2}
\end{equation}
The state vector in the interaction picture $|\psi(t)\ra_I$ obeys the 
equation of motion
\begin{eqnarray}
	 i{d\over{dt}}|\psi(t)\ra_I&=& {\cal H}_{B1}^\prime(t)|\psi(t)\ra_I,
	\nonumber  \\
	{\cal H}_{B1}^\prime(t) &=&e^{iH_{B0}^\prime t} H_{B1}^\prime 
	e^{-iH_{B0}^\prime t}=\mu(t)e^{i\epsilon t}\,J_+
+\mu(t)^*e^{-i\epsilon t}\,J_-.
	\label{binint}
\end{eqnarray}
Let us suppose that the system is in the lowest weight state 
$|M,0\ra$ at $t=0$. 
Then we obtain
\begin{eqnarray}
    	|\psi(t)\ra_I& =& T\,\exp\left[-i\int_0^t{\cal H}_{B1}^\prime
    	(t^\prime)dt^\prime \right]
	|M,0\ra, \nonumber\\
	& =& T\exp\left[-i\int_0^t\mu(t^\prime)
        e^{i\epsilon t^\prime}dt^\prime 
	J_+
	-i\int_0^t\mu(t^\prime)^*e^{-i\epsilon t^\prime}dt^\prime 
	J_-\right]|M,0\ra.
	\label{bingensol}
\end{eqnarray}
Since $g=T\,e^{-i\int_0^t{\cal H}_{B1}^\prime(t^\prime)dt^\prime}\in 
SU(2)$ it can always be decomposed into $g=\exp(\zeta 
J_+-\zeta^*J_-) \exp(i\nu J_0)$ ($\zeta\in {\bf C}, \nu\in{\bf R}$) 
and the obtained state is the binomial 
state.
For illustration purpose let us choose a special form of $\mu(t)$:
\begin{equation}
	\mu(t)=i e^{-i\epsilon t+i\theta}\eta, \quad \eta\in {\bf R}.
	\label{muchoice}
\end{equation}
Then we obtain \cite{lee,fus4}
\begin{eqnarray}
|\psi(t)\ra_I &=&\exp\left[e^{i\theta}\eta t\,J_+-e^{i\theta}\eta 
t\,J_-\right]|M,0\ra,\nonumber\\
              &\propto& \exp\left[e^{i\theta}\tan(\eta 
              t)\,J_+\right]|M,0\ra \propto|\tan(\eta t)e^{i\theta};M\ra.
              \label{bingensol2}
\end{eqnarray}
              
\bigbreak
However this is not exactly the same as the binomial state 
(\ref{disbin}) given in the previous Appendix.
In order to relate these two forms let us note that the state in the 
Fock space  (\ref{binfock}) is uniquely specified by $n_1\equiv n$ 
only:
\begin{equation}
	\bb n\ra\equiv|M-n,n\ra.
	\label{twofock}
\end{equation}
Let us understand that the states $\{\bb n\ra$, $n=0,1,\ldots,M\}$ are 
generated by the ``reduced'' single boson operators $b^\dagger$ and $b$
as in (\ref{oscil2}).
Then the $su(2)$ operators are expressed in terms of $b^\dagger$ and 
$b$ as
\begin{eqnarray}
J_+\bb n\ra&=&J_+|M-n,n\ra=a_1^\dagger a_0|M-n,n\ra, \nonumber\\           
           &=&\sqrt{n+1}\sqrt{M-n}\,|M-n-1,n+1\ra=\sqrt{n+1}\sqrt{M-n}\,\bb 
           n+1\ra, \nonumber\\
           &=& b^\dagger\sqrt{M-N}\,\bb n\ra,\qquad \ N=b^\dagger b,
           \label{hprel}
\end{eqnarray}           
namely
\begin{equation}
	{\cal J}_+=b^\dagger\sqrt{M-N},\quad {\cal J}_-=\sqrt{M-N}\,b,\quad
	{\cal J}_0=N-{M\over2}.
	\label{hpsu2}
\end{equation}
Thus we have naturally ``derived'' the H-P realisation of $su(2)$ used in the 
previous Appendix. The lowest weight state in this notation is
$$
\bb0\ra\equiv|M,0\ra,\quad {\cal J}_-\bb0\ra=0,\quad {\cal 
J}_0\bb0\ra=-{M\over2}\bb0\ra.
$$

\bigbreak
\bigbreak
At the end of the previous Appendix we have shown that the binomial
state tends to the ordinary coherent state in a certain limit.
Here we will show a result in an opposite direction.
That is, the  binomial states can be obtained from the ordinary coherent
states with two degrees of freedom by appropriate `slicing' or 
restriction.
This reveals some features of the binomial states quite naturally.
As before let us start with the corresponding result in the probability 
theory, which is rather elementary.
A double Poisson distribution is given by
\begin{equation}
	P_{n_0,n_1}(\alpha_0,\alpha_1)\equiv P_{\bf n}
	(\hbox{\boldmath $\alpha$})=
	e^{-\alpha_0^2-\alpha_1^2}\,
	{\alpha_0^{2n_0}\alpha_1^{2n_1}\over{n_0!n_1!}},\quad {\bf 
	n}=(n_0,n_1),\quad \hbox{\boldmath $\alpha$}=(\alpha_0,\alpha_1).
	\label{doubpoi}
\end{equation}
If we restrict it to a line
$$n_0+n_1=M,
$$
we obtain the binomial distribution up to normalisation:
\begin{eqnarray}
P_{M-n,n}(\alpha_0,\alpha_1)&=&e^{-\alpha_0^2-\alpha_1^2}
{\alpha_0^{2(M-n)}\alpha_1^{2n}\over{(M-n)!n!}},\qquad \ 
\eta\equiv{\alpha_1\over\sqrt{\alpha_0^2+\alpha_1^2}}.\nonumber\\
&\propto&{M\choose n}\eta^{2n}(1-\eta^2)^{M-n}=B_n(\eta ;M).
\label{poitobin}
\end{eqnarray}  

The same proposition at the level of the ``probability amplitude''
including the normalisation can be easily obtained by considering the 
projection operator onto the representation space 
(\ref{binfock}) \cite{FKSF}:
\begin{equation}
	{\bf P}_M=\sum_{n_0+n_1=M}|n_0,n_1\ra\la n_0,n_1|.
	\label{projM}
\end{equation}
With the aid of the resolution of unity (over-completeness relation)
\begin{equation}
    \int{d^2\alpha_{0C}d^2\alpha_{1C}\over{\pi^2}}
	|\hbox{\boldmath $\alpha$}_C\ra\la\hbox{\boldmath $\alpha$}_C|=1,\quad 
	\hbox{\boldmath $\alpha$}_C=(\alpha_{0C},\alpha_{1C}), \quad 
	d^2\alpha_{jC}=d(\alpha_{jC})_Rd(\alpha_{jC})_I,\quad j=0,1,
	\label{overcomp}
\end{equation}
for the double coherent state
$$
|\hbox{\boldmath $\alpha$}_C\ra=e^{-|\hbox{\boldmath $\alpha$}_C|^2/2}
\sum_{n_0,n_1}{(\alpha_{0C})^{n_0}(\alpha_{1C})^{n_1}\over
\sqrt{n_0!n_1!}}|n_0,n_1\ra
$$
we have
\begin{equation}
    {\bf P}_M= \int{d^2\alpha_{0C}d^2\alpha_{1C}\over{\pi^2}}
    |\hbox{\boldmath $\alpha$}_C\ra\la\hbox{\boldmath $\alpha$}_C|\,{\bf P}_M.
		\label{projiden}
\end{equation}
By a change of variables
\begin{equation}
	{\alpha_{0C} \choose 
	\alpha_{1C}}={\zeta_C\over\sqrt{1+|\xi_C|^2}}{1\choose \xi_C}
	\label{chvar}
\end{equation}
we have
\begin{eqnarray}
	{\rm r.h.s.\ of\ (\ref{projiden})} & = & 
	\int{d^2\xi_C\over{\pi(1+|\xi_C|^2)^2}}
	\int{|\zeta_C|^2d^2\zeta_C\over\pi}e^{-|\zeta_C|^2}\non \\
	& &\times\sum_{m_0,m_1=0}^\infty
	{1\over\sqrt{m_0!m_1!}}
        \left({\zeta_C\over\sqrt{1+|\xi_C|^2}}\right)^{m_0+m_1}
	\xi_C^{m_1}|m_0,m_1\ra
	\non  \\
	 &  & 
	 \times\sum_{n_0+n_1=M}\la n_0,n_1|{1\over\sqrt{n_0!n_1!}}
	 \left({\zeta_C^*\over\sqrt{1+|\xi_C|^2}}\right)^M
	\xi_C^{*n_1}
	\non  \\
	 & = & \int{d^2\xi_C\over\pi}{(M+1)!\over{(1+|\xi_C|^2)^{M+2}}}
	 \sum_{m_0+m_1=M}
	{1\over\sqrt{m_0!m_1!}}\xi_C^{m_1}|m_0,m_1\ra
	\non  \\
	 &  & \times\sum_{n_0+n_1=M}\la n_0,n_1|{1\over\sqrt{n_0!n_1!}}
	\xi_C^{*n_1}
	\non\\
		 & = & \int d\mu(\xi_C,\xi_C^*)|\xi_C\ra\la\xi_C|,
	\label{binmeas}
\end{eqnarray}
in which
\begin{eqnarray}
	|\xi_C\ra & = & {1\over{(1+|\xi_C|^2)^{M/2}}}
        \sum_{n=0}^M\sqrt{M \choose 
	n}\,\xi_C^n|M-n,n\ra
	\non  \\
	 & = &  {1\over{(1+|\xi_C|^2)^{M/2}}}\sum_{n=0}^M\sqrt{M \choose 
	n}\,\xi_C^n\bb n\ra
	\label{anotherbin}
\end{eqnarray}
and the measure is
\begin{equation}
	d\mu(\xi_C,\xi_C^*)={(M+1)!\over{ 
	M!}}{d^2\xi_C\over{\pi(1+|\xi_C|^2)^2}}.
	\label{binmeaasform}
\end{equation}
By introducing a  parameter $\eta_C\equiv{\xi_C/\sqrt{1+|\xi_C|^2}}$, 
we can identify $|\xi_C\ra$ as the binomial state $|\eta_C ;M\ra$ 
(\ref{binst}). This process shows  elementarily that the 
parameter space of the binomial states is $SU(2)/U(1)={\bf CP}^1$ 
($\xi_C=\alpha_{1C}/\alpha_{0C}$) obtained from ${\bf C}^2$ 
({\boldmath $\alpha_C$}=($\alpha_{0C},\alpha_{1C}$)) by integrating 
out the overall factor $\zeta_C$.

\section*{Appendix C\ \ Multinomial States}
\setcounter{equation}{0}
\renewcommand{\theequation}{C.\arabic{equation}} 
The multinomial distribution is
\begin{equation}
	M_{\bf n}(\hbox{\boldmath $\eta$}\,;M)={M!\over{{\bf 
	n}!}}\eta_1^{2n_1}\cdots\eta_r^{2n_r}(1-\hbox{\boldmath 
	$\eta$}^2)^{n_0},
	\label{mulnomdis}
\end{equation}
in which
\begin{eqnarray}
	{\bf n}&=&(n_0,n_1,\ldots,n_r),\quad n_0+n_1+\cdots+n_r=M,
	\quad \hbox{\boldmath $\eta$}=(\eta_1,\ldots,\eta_r)\in{\bf 
	R}^r,\non\\
	 0&<&\hbox{\boldmath $\eta$}^2=\eta_1^2+\cdots+\eta_r^2<1,
	\quad {\bf n}!=n_0!n_1!\cdots n_r!.
	\label{mulnot}
\end{eqnarray}
Let us first define the `{\it multinomial state}' in the linear 
representation form
\begin{equation}
	|\hbox{\boldmath $\eta$}_C\,;M\ra=\sum_{\bf n}
	      \sqrt{M!\over{{\bf n}!}}(\eta_{1C})^{n_1}\cdots(\eta_{rC})^{n_r}
	      (1-|\hbox{\boldmath $\eta$}_C|^2)^{n_0/2}|{\bf n}\ra,
	\label{multist}
\end{equation}
in which the Fock states
\begin{equation}
	|{\bf n}\ra=|n_0,n_1,\ldots,n_r\ra,\quad   n_0+n_1+\cdots+n_r=M,
	\label{mulfock}
\end{equation}
are generated by $r+1$ bosonic oscillators
\begin{eqnarray}
	[a_j,a_k^\dagger]&=&\delta_{jk},\quad a_j|{\bf 0}\ra=0,
	\quad j=0,1,\ldots,r, \quad	|{\bf 0}\ra=|0,0,\ldots,0\ra,\non\\ 
	|{\bf n}\ra&=&{({\bf a}^\dagger)^{\bf n}\over\sqrt{{\bf n}!}}|{\bf 
	0}\ra,\quad 
	({\bf a}^\dagger)^{\bf 
	n}={a_0^\dagger}^{n_0}{a_1^\dagger}^{n_1}\cdots{a_r^\dagger}^{n_r}.
	\label{mulosci}
\end{eqnarray}
Obviously the above Fock space (\ref{mulfock}) provides an irreducible 
representation of $su(r+1)$ with generators
\begin{equation}
	J_{jk}=a_j^\dagger a_k,\quad j\neq k,\quad N_j=a_j^\dagger a_j,
	\label{sur1gen}
\end{equation}
in which $J_{jk}$ ($j>k$) are considered as shift-up operators.
It is a symmetric representation corresponding to the same Young 
diagram as before:
\beq
\square\square\square \cdots \square\square\ \ M\ {\rm boxes},
\label{suryoung}
\eeq 
and the lowest weight state is
\begin{equation}
	|{\bf 0}^\prime\ra\equiv|M,0,\ldots,0\ra,\quad J_{0k}|{\bf 
	0}^\prime\ra=0,\quad 
	J_{jk}|{\bf 0}^\prime\ra=0,\quad j,\,k>0.
	\label{mullowwei}
\end{equation}
The last equation shows that the lowest weight state 
$|{\bf 0}^\prime\ra$ is invariant under $u(r)$.
The dimension of the above irreducible representation is
\begin{equation}
	{M+r \choose M}={M+r \choose r}
	\label{symirrepdim}
\end{equation}
which is the same as the number of terms in the multinomial expansion,
the number of the partitions of $M$ into $r+1$ non-negative integers
and the number of $M$-th order partial derivatives of analytic 
functions of $r+1$ variables.
It should be remarked that there are other types of coherent states 
of $su(r+1)$ ($r\ge2$) algebra belonging to the Young diagrams  other than 
those given above (\ref{suryoung}). They cannot be constructed by bosons only.

\bigbreak
It is not difficult to generate the multinomial states explicitly 
by applying the $SU(r+1)$ operator on the lowest weight (energy) state 
$|{\bf 0}^\prime\ra$.
For simplicity and concreteness let us show this for the $su(3)$ case:
\begin{equation}
	|\eta_1e^{i\theta_1}, \eta_2e^{i\theta_2}\,;M\ra=\sum_{n_1,n_2=0}^M
	\sqrt{M!\over{n_0!n_1!n_2!}}(\eta_1e^{i\theta_1})^{n_1}
        (\eta_2e^{i\theta_2})^{n_2}
	(1-\eta_1^2-\eta_2^2)^{n_0/2}|n_0,n_1,n_2\ra.
	\label{trinomst}
\end{equation}
This process is essentially the same as the generation of negative 
trinomial state given in section 4.
For the `{\it trinomial state}' we only have to use the $su(2)$ 
disentangling theorems twice for two $su(2)$ subalgebras spanned by 
$(0,1)$ and $(1,2)$ oscillators:
\begin{eqnarray}
	& &e^{r(e^{i\theta_1}J_{10}-e^{-i\theta_1}J_{01})}  
	\label{disen1}  \\
	 & = & \exp[e^{i\theta_1}\tan 
	 r\,J_{10}]\exp[\log(1+\tan^2r)(N_1+N_2/2-M/2)]
         \exp[-e^{-i\theta_1}\tan 
	 r\,J_{01}],
	\non\\
         & &e^{r^\prime(e^{i\theta_2}J_{21}-e^{-i\theta_2}J_{12})} 
	\label{disen2} \\
	 & = & \exp[e^{i\theta_2}\tan 
	 r^\prime\,J_{21}]\exp[-{1\over2}\log(1+\tan^2r^\prime)
        (N_1-N_2)]\exp[-e^{-i\theta_2}\tan 
	 r^\prime\,J_{12}].
	\non	
\end{eqnarray}
We choose $r$  and $r^\prime$ such that
$$
\tan^2r={\eta_1^2+\eta_2^2\over{1-\eta_1^2-\eta_2^2}},
\quad \tan^2r^\prime={\eta_2^2\over{\eta_1^2}}.
$$
The generalisation to higher rank cases is straightforward.
One has to apply $su(2)$ disentangling theorems in the following 
sequence of $su(2)$ algebras spanned by
$(0,1)$, $(1,2)$, $(2,3)$, $\ldots$, $(r-1,r)$ oscillators.

\bigbreak
\bigbreak
To obtain the $su(r+1)$ multinomial states from the $R+1$-fold
coherent states is also straightforward. One only needs to develop 
clever notation to express the essential features succinctly.

A multiple Poisson distribution is given by
\begin{equation}
	 P_{\bf n}
	(\hbox{\boldmath $\alpha$})=
	e^{-\alpha_0^2-\cdots-\alpha_r^2}\,	
	{\alpha_0^{2n_0}\alpha_1^{2n_1}\cdots\alpha_r^{2n_r}
        \over{n_0!\cdots\,n_r!}},\quad {\bf 
	n}=(n_0,n_1,\ldots,n_r),\quad \hbox{\boldmath 
	$\alpha$}=(\alpha_0,\alpha_1,\ldots,\alpha_r).
	\label{mulpoi}
\end{equation}
If we restrict it to a hyperplane
$$n_0+n_1+\cdots+n_r=M,\quad \hbox{or}\quad n_0=M-n_1-\cdots-n_r,
$$
we obtain the multinomial distribution up to normalisation :
\begin{eqnarray}
\eta_j&\equiv&{\alpha_j\over\sqrt{\hbox{\boldmath$\alpha$}^2}},\quad
\hbox{\boldmath$\alpha$}^2=\sum_{j=0}^r\alpha_j^2,\nonumber\\
P_{\bf n}(\hbox{\boldmath$\alpha$})
&\propto&{M!\over{{\bf n}!}}\eta_1^{2n_1}\cdots\eta_r^{2n_r}
(1-\hbox{\boldmath$\eta$}^2)^{n_0}=M_{\bf n}(\hbox{\boldmath$\eta$} ;M).
\label{mulpoitomul}
\end{eqnarray}  

The same proposition at the level of the ``probability amplitude''
including the normalisation can be easily obtained by considering the 
projection operator onto the representation space (\ref{mulfock})
 \cite{FKSF}:
\begin{equation}
	{\bf P}_M=\sum_{n_0+n_1+\cdots+n_r=M}
	|n_0,n_1,\ldots,n_r\ra\la n_0,n_1,\ldots,n_r|.
	\label{projM2}
\end{equation}
With the aid of the resolution of unity (over-completeness relation)
\begin{eqnarray}
    \int{\prod_{j=0}^rd^2\alpha_{jC}\over{\pi^{r+1}}}
	|\hbox{\boldmath $\alpha$}_C\ra\la\hbox{\boldmath $\alpha$}_C|
        &=&1,\qquad 
	\hbox{\boldmath $\alpha$}_C
	=(\alpha_{0C},\alpha_{1C},\ldots,\alpha_{rC}),\non\\ 
	d^2\alpha_{jC}&=&d(\alpha_{jC})_Rd(\alpha_{jC})_I,\quad j=0,\ldots,r
	\label{overcomp2}
\end{eqnarray}
for the multiple coherent states
$$
|\hbox{\boldmath $\alpha$}_C\ra=e^{-|\hbox{\boldmath $\alpha$}_C|^2/2}
\sum_{n_0,\ldots,n_r}{(\alpha_{0C})^{n_0}\cdots(\alpha_{rC})^{n_r}\over
\sqrt{{\bf n}!}}|n_0,n_1,\ldots,n_r\ra
$$
we have
\begin{equation}
    {\bf P}_M= \int{\prod_{j=0}^rd^2\alpha_{jC}\over{\pi^{r+1}}}
    |\hbox{\boldmath $\alpha$}_C\ra\la
    \hbox{\boldmath $\alpha$}_C|\,{\bf P}_M.
		\label{projiden2}
\end{equation}
By a change of variables
\begin{equation}
	\pmatrix{\alpha_{0C} \cr \alpha_{1C}\cr \vdots\cr \alpha_{rC}\cr}=
	{\zeta_C\over\sqrt{1+|\hbox{\boldmath$\xi$}_C|^2}}\pmatrix{1\cr 
	\xi_{1C}\cr \vdots \cr \xi_{rC}\cr}
	\label{chvar2}
\end{equation}
we have
\begin{eqnarray}
{\bf P}_M & = & 
\int{\prod_{j=1}^rd^2\xi_{jC}\over{\pi^r(1+|\hbox{\boldmath$\xi$}_C|^2)^2}}
\int{|\zeta_C|^{2r}d^2\zeta_C\over\pi}e^{-|\zeta_C|^2}\non \\
& &\times\sum_{m_0,m_1,\ldots,m_r=0}^\infty
{1\over\sqrt{\bf m!}}\left({\zeta_C\over\sqrt{1+|\hbox{\boldmath$\xi$}_C|^2}}
\right)^{m_0+m_1+\cdots+m_r}
\xi_{1C}^{m_1}\xi_{2C}^{m_2}\cdots\xi_{rC}^{m_r}|m_0,m_1,\ldots,m_r\ra
\non  \\
 &  & 
 \times\sum_{n_0+n_1+\ldots+n_r=M}\la 
 n_0,n_1,\ldots,n_r|{1\over\sqrt{\bf n!}}
 \left({\zeta_C^*\over\sqrt{1+|\hbox{\boldmath$\xi$}_C|^2}}\right)^M
\xi_{1C}^{*n_1}\xi_{2C}^{*n_2}\cdots\xi_{rC}^{*n_r}
\non  \\
 & = & \int{\prod_{j=1}^rd^2\xi_{jC}\over\pi^r}{(M+r)!\over
 {(1+|\hbox{\boldmath$\xi$}_C|^2)^{M+r+1}}}
 \sum_{m_0+m_1+\ldots+m_r=M}
{1\over\sqrt{\bf m!}}\xi_{1C}^{m_1}\xi_{2C}^{m_2}
\cdots\xi_{rC}^{m_r}|m_0,m_1,\ldots,m_r\ra
\non  \\
 &  & \times\sum_{n_0+n_1+\ldots+n_r=M}\la 
 n_0,n_1,\ldots,n_r|{1\over\sqrt{\bf n!}}\xi_{1C}^{*n_1}
\xi_{2C}^{*n_2}\cdots\xi_{rC}^{*n_r}
\non\\
 & = & \int d\mu(\hbox{\boldmath$\xi$}_C,\hbox{\boldmath$\xi$}_C^*)
 |\hbox{\boldmath$\xi$}_C\ra\la\hbox{\boldmath$\xi$}_C|,
\label{mulmeas}
\end{eqnarray}
in which
\begin{eqnarray}
|\hbox{\boldmath$\xi$}_C\ra & = & 
{1\over{(1+|\hbox{\boldmath$\xi$}_C|^2)^{M/2}}}\sum_{\bf n}\sqrt{M! 
\over{ {\bf 
n}!}}\,\xi_{1C}^{n_1}\xi_{2C}^{n_2}\cdots\xi_{rC}^{n_r}|M-\sum 
n_j^\prime,{\bf n^\prime}\ra
\non  \\
 & = &  {1\over{(1+|\hbox{\boldmath$\xi$}_C|^2)^{M/2}}}\sum_{\bf n}
 \sqrt{M! \over{{\bf n}!}}\,\xi_{2C}^{n_2}\cdots\xi_{rC}^{n_r}\bb {\bf 
 n^\prime}\ra
\label{anothermul}
\end{eqnarray}
and the measure is
\begin{equation}
d\mu(\hbox{\boldmath$\xi$}_C,\hbox{\boldmath$\xi$}_C^*)={(M+r)!\over{ M!}}
{\prod_{j=1}^rd^2\xi_{jC}\over{\pi^r(1+|\hbox{\boldmath$\xi$}_C|^2)^{r+1}}}.
\label{mulmeasform}
\end{equation}
By introducing   parameters 
$\eta_{jC}\equiv{\xi_{jC}/\sqrt{1+|\hbox{\boldmath$\xi$}_C|^2}}$, 
we can identify $|\hbox{\boldmath$\xi$}_C\ra$ as the multinomial state 
$|\hbox{\boldmath$\eta$}_C ;M\ra$ 
(\ref{multist}). This process shows  elementarily that the 
parameter space of the multinomial states is 
$SU(r+1)/U(1)\times SU(r)={\bf CP}^r$ 
($\xi_{jC}=\alpha_{jC}/\alpha_{0C}$) obtained from ${\bf C}^{r+1}$ 
({\boldmath $\alpha_C$}) by integrating 
out the overall factor $\zeta_C$.

\bigbreak
A few words about the multiple coherent states limit of the 
multinomial states.
For the multinomial state (\ref{anothermul}) we let $M\to\infty$ and 
$\xi_{jC}\to0$ while keeping the `average' 
fixed, $\xi_{jC}^2M=\alpha_{jC}^2$ to obtain for fixed ${\bf 
n}^\prime$
\begin{equation}
	|\hbox{\boldmath$\xi$}_C\ra \to
	 e^{-|\hbox{\boldmath$\alpha$}_C|^2/2}\sum_{\bf 
	 n^\prime}{(\alpha_C)^{\bf n^\prime}\over{{\bf n^\prime}!}}\,\bb {\bf 
	 n^\prime}\ra.
	\label{multtonucohe}
\end{equation}

\bigbreak
\bigbreak
Like in the case of the binomial states one can express the states and 
the $su(r+1)$ generators in the ``reduced'' notation using only $r$ 
boson oscillators.
This gives rise to the generalisation of the Holstein-Primakoff
realisation.
But as remarked above it is applicable only to the {\it symmetric 
representations}. 
Because of the constraint
$$n_0+n_1+\cdots+n_r=M,
$$
the state $|{\bf n}\ra$ is uniquely specified by
$$ {\bf n}^\prime=(n_1,n_2,\ldots,n_r)$$
only. So we identify
\begin{equation}
	\bb{\bf n}^\prime\ra=\bb 
	n_1,n_2,\ldots,n_r\ra\equiv|M-\sum^\prime n_j,n_1,\ldots,n_r\ra=|{\bf 
	n}\ra,
	\label{redfock}
\end{equation}
and introduce $r$ independent boson oscillators
$$
[b_j,b_k^\dagger]=\delta_{jk}, \quad b_j\bb 0\ra=0,\quad j=1,\ldots,r,
$$
which create the ``reduced'' states
\begin{equation}
    \bb{\bf n^\prime}\ra={({\bf b}^\dagger)^{\bf n^\prime}\over{{\bf 
	n^\prime}!}}\bb{\bf 0}\ra,\quad 
		({\bf b}^\dagger)^{\bf 
		n^\prime}={b_1^\dagger}^{n_1}\cdots{b_r^\dagger}^{n_r},\quad
		{\bf n^\prime}!=n_0!\cdots n_r!.
     \label{muloscib}
\end{equation}
Then we have
\begin{equation}
	{\cal J}_{j0}=b_j^\dagger\sqrt{M-N_1-\cdots-N_r},\quad 	
	{\cal J}_{0j}=\sqrt{M-N_1-\cdots-N_r}\,b_j,\quad {\cal 
	J}_{jk}=b_j^\dagger b_k.
	\label{hpmul}
\end{equation}
Note that the ``vacuum'' $\bb 0\ra$ is the lowest weight state and it 
is invariant under $su(r)$ which is expressed {\it linearly}:
\begin{equation}
	{\cal J}_{0j}\bb0\ra=0,\quad {\cal J}_{jk}\bb0\ra=0.
	\label{hpmulvac}
\end{equation}

\bigbreak
Before closing this Appendix, let us remark on the dynamical 
generation of the multinomial states. This is essentially the same as
that of the binomial states.
Let us consider a collection (total number $M$) of identical 
$r+1$-level atoms (bosons). It is assumed that the interactions among
different atoms are negligibly small compared with the interactions 
within the same atoms among different energy levels.
As before the system is described in terms of $r+1$ bosonic oscillators
and the Hamiltonian at the zero-th order approximation is quadratic in 
the oscillators keeping the total number of atoms fixed.
In other words the Hamiltonian is a hermitian linear combination of
the $u(r+1)$ generators given in (\ref{sur1gen}).
If we assume that the system is in the lowest energy (weight) state
$|M,0,\ldots,0\ra$ at $t=0$, then at time $t$ it is
$$
e^{-iHt}|M,0,\ldots,0\ra,
$$
which is a multinomial state since the time evolution operator 
$e^{-iHt}$ is an element of $U(r+1)$ and the $U(1)$ part and the 
$SU(r)$ is immaterial when they act on $|M,0,\ldots,0\ra$.


\section*{Appendix D\ \ Waiting Time : Negative Binomial Distribution}
\setcounter{equation}{0}
\renewcommand{\theequation}{D.\arabic{equation}} 
For those who are not familiar with probability theory, we give here
a simple example in which the {\it negative binomial distribution} 
occurs.
We follow Feller's textbook \cite{Feller}. Let us consider a 
succession of Bernoulli's trials each of which has the probability of 
{\it failure} $0<\eta^2<1$. We ask a question: How long it will take
for the $M$-th success to turn up? Here $M$ is a positive integer.
Since $M$-th success comes not earlier than $M$-th try, we denote by
$B_n^-(\eta ;M)$ the probability that the $M$-th success occurs at 
the trial number $M+n$, $n\ge0$.
This occurs, if and only if, among the $M+n-1$ trials there are exactly $n$
failures and the $M+n$-th trial results in success: so that
$$
B_n^-(\eta ;M)={M+n-1 \choose n}(1-\eta^2)^M\eta^{2n}.
$$
For the very unlucky the waiting time ($n$) can be infinite.
This corresponds to the fact that the irreducible unitary 
representations of non-compact algebras are infinite dimensional.
 

\end{document}